\documentclass[twocolumn]{openjournal}
\usepackage{graphicx} % Required for inserting images

\usepackage[colorlinks,linkcolor=blue,citecolor=blue,urlcolor=blue ]{hyperref}
\usepackage[utf8]{inputenc}
\usepackage{float}
\usepackage{xcolor}
\usepackage{ulem}
\usepackage[T1]{fontenc}

\newcommand{\msolar} {$\rm{M_{\odot}}~$}
\newcommand{\msolarc} {$\rm{M_{\odot}}$}
\newcommand{\enzo}{\texttt{Enzo~}}

\newcommand{\renaissance} {\texttt{Renaissance~}}
\newcommand{\LCDM}{$\Lambda$CDM}

\def\jwadd#1{{\color{black} #1}}

\begin{document}
\title{No Tension: JWST Galaxies at $\MakeLowercase{z} >10$ Consistent with Cosmological Simulations\vspace{-3em}}
\author{Joe M. McCaffrey$^{1,*}$}
\author{Samantha E. Hardin$^2$}
\author{John H. Wise$^2$}
\author{John A. Regan$^1$}
\thanks{$^*$E-mail:joe.mccaffrey@mu.ie}
\affiliation{$^1$Centre for Astrophysics and Space Science Maynooth, Department of Theoretical Physics, Maynooth University, Maynooth, Ireland \\
$^2$Center for Relativistic Astrophysics, Georgia Institute of Technology, 837 State Street, Atlanta, GA 30332, USA}\begin{abstract}
    {\color{black}\noindent Recent observations by JWST have uncovered galaxies in the very early Universe via the JADES and CEERS surveys. These galaxies have been measured to have very high stellar masses with substantial star formation rates. There are concerns that these observations are in tension with the $\Lambda$CDM model of the Universe, as the stellar masses of the galaxies are relatively high for their respective redshifts. Recent studies have compared the JWST observations with large-scale cosmological simulations. While they were successful in reproducing the galaxies seen in JADES and CEERS, the mass and spatial resolution of these simulations were insufficient to fully capture the early assembly history of the simulated galaxies. In this study, we use results from the \renaissance simulations, which are a suite of high resolution simulations designed to model galaxy formation in the early Universe.
    We find that the most massive galaxies in \renaissance have stellar masses and star formation rates that are consistent with the observations from the JADES and CEERS surveys. The \jwadd{higher} resolution afforded by \renaissance allows us to model the build-up of early galaxies from stellar masses as low as $10^4$ M$_{\odot}$ up to a maximum stellar mass of a few times $10^7$ M$_{\odot}$. Within this galaxy formation paradigm, {\color{black}and after extrapolating forward in time the stellar masses in Renaissance where required}, we find {\color{black}overall} agreement with JADES and CEERS.
    We find no tension between the $\Lambda$CDM model and current JWST measurements. As JWST continues to explore the high redshift Universe, simulations, such as \texttt{Renaissance}, will continue to be crucial in understanding the formation history of early embryonic galaxies.}\\[1em]
\end{abstract}
\maketitle

\section{Introduction}
\noindent With the launch, and now the first observations with JWST, the high-redshift Universe is being unveiled with unprecedented detail for the first time. 
As exquisitely detailed measurements of distant early galaxies are now within range of observations, it allows for the opportunity to {\color{black}compare} the results of these observations against high resolution simulations of early galaxy formation which was previously intractable.\par
 The JADES survey \citep{Bunker_2019} has provided measurements on five galaxies with spectroscopically confirmed redshifts at $z>10$. Three of these galaxies are the most distant yet detected.  \cite{robertson}, \cite{curtis} and \cite{bunker2023jades} have constrained physical properties of these five galaxies, finding that the galaxies lie at (mean) spectroscopic redshifts of $z = 10.38$ (GS-z10-0), $z = 11.58$ (GS-z11-0), $z = 12.63$ (GS-z12-0), $z = 13.20$ (GS-z13-0) and $z = 10.60$ (GN-z11).
 Additionally the CEERS project \citep{CEERS} has also provided measurements for Maisie's galaxy which has a redshift of 11.44 \citep{Maisie}. In total there are six spectroscopically confirmed galaxies against which we can directly compare.\\
 \indent The JADES survey, performed using the NIRCam instrument on JWST, targets a region, previously studied by the Hubble Telescope \citep{HUDF}, in nine different wavelength ranges. JADES was conducted with the aim of detecting faint galaxies using the dropout-technique \citep{Bunker_2019} allowing for fast identification of high redshift galaxy candidates. However, the photometry alone is not enough to confirm the candidates' redshift and a follow-up spectrum, using an instrument like NIRSpec, is needed to confidently quantify the redshift of the candidates as noted above.\\
\indent Similar to JADES, CEERS aims to study the first 500 Myr of galaxy evolution again by using a combination of the NIRCam instrument for fast identification followed by a longer duration follow-up using NIRSpec. Initial photometric measurements of Maisie's galaxy and of CEERS-93316 placed them at photometric redshifts of $z_{\rm{phot}} = 11.08$ \citep{Maisie_initial} and $z_{\rm{phot}} = 16.45$ \citep{CEERS-93316-initial}, respectively.  Spectroscopic follow-up measurements of these galaxies confirmed their redshifts to be $z_{\rm spec} = 11.44$ and $z_{\rm spec} = 4.912$ \citep{Maisie}, respectively. Due to the decrease in redshift of CEERS-93316 to a much lower redshift, we do not include it in the analysis in this paper and instead only include the high redshift Maisie's galaxy.\\
\indent The spectra of the galaxies found in JADES and CEERS were analysed using \texttt{Beagle} \citep{beagle}, to estimate the stellar mass and star formation rate of each galaxy which we compare directly against our simulations. More details on the modelling procedure used to reduce the observational data can be found in the detection papers \citep[e.g.][]{bunker2023jades, Maisie}.

\par In a recent study \citep[][hereafter K23]{Keller_2023} tested the capabilities of a variety of cosmological simulations by investigating whether these simulations were able to reproduce galaxies with properties similar to the galaxies observed in the JADES and CEERS surveys. 
To do this, K23 utilised EAGLE \citep{EAGLE}, Illustris \citep{Illustris}, TNG100 \citep{TNG100}, RomulusC \citep{RomulusC}, {\sc Obelisk} \citep{Obelisk} and Simba \citep{simba}.
\par K23 concluded that the cosmological simulations they examined were able to reproduce galaxies with a similar stellar mass and star formation rate (SFR) compared to the galaxies observed in the JADES simulations and that the observed galaxies are consistent with a flat $\Lambda$CDM model. In this paper, we build-on the work of K23 by comparing the JADES and CEERS results against high resolution simulations that were designed to specifically examine a high-redshift environment and to model the early assembly history of the first galaxies. {\color{black}By considering simulation data from both the higher halo mass regime of K23 and the lower halo mass data from this paper, we strengthen the argument that there exists no tension between the measurements of JWST and the $\Lambda$CDM model of the Universe.}
 
\par This study accomplishes this using the \texttt{Renaissance} simulation suite \citep{renaissance, chen_renaissance, OShea_2015}. The \texttt{Renaissance} simulations model early galaxy formation in three regions which differ from each other by their level of overdensity \citep[see][for details]{OShea_2015}. Using a similar methodology to K23, we compare the results from these  simulations to the JADES and CEERS galaxy property estimates, observationally validating our simulation results and also determining the likeliness of such massive galaxies forming early in a $\Lambda$CDM cosmology.  \\
\indent There have been significant concerns that the early measurements by JWST are in conflict with the $\Lambda$CDM model of the Universe. In particular, there are claims that the stellar masses of the galaxies observed by JWST are simply too {\color{black}high} to be explained {\color{black}in the context of} a $\Lambda$CDM cosmology \citep[e.g.][]{Haslbauer_2022} and that the masses of the JWST galaxies, as measured at redshifts between 7 and 10 by \cite{Labbe_2022} in particular, are testing the upper limits on the available baryonic mass available according to $\Lambda$CDM \citep{boylankolchin2023stress}. See also \citet{Lovell-tension}, \citet{Dekel-tension}, \citet{Harikane-2023-tension}, \citet{Steinhardt-tension}, \citet{Menci-tension}, and \citet{Mason-tension} for more discussion and analysis on the proposed tension between the JWST measurements and $\Lambda$CDM.\par
\indent This paper addresses these concerns by showing that simulations (based on a $\Lambda$CDM cosmology) are able to reproduce galaxies consistent with the early findings of JWST at least at the very highest redshifts explored by it. If the \cite{Labbe_2022} result holds under spectroscopic scrutiny and the galaxies are found to lie in typical regions of a $\Lambda$CDM Universe, then, as pointed out by \cite{boylankolchin2023stress}, this represents a major challenge to standard cosmology. However, see also \cite{Prada_2023} which offers an alternative explanation to the large stellar masses found by \cite{Labbe_2022}.\par
\indent The paper is laid out as follows: \S \ref{method} describes the high-resolution \texttt{Renaissance} simulations as well as the methodology and code used to run the suite, \S \ref{results} discusses the results of the analysis performed and in \S \ref{conlusions} we discuss the implications of the results and the case for further analysis of the high redshift Universe using cosmological simulations.
% \newpage

\begin{figure}
    \centering
    \includegraphics[width=\columnwidth]{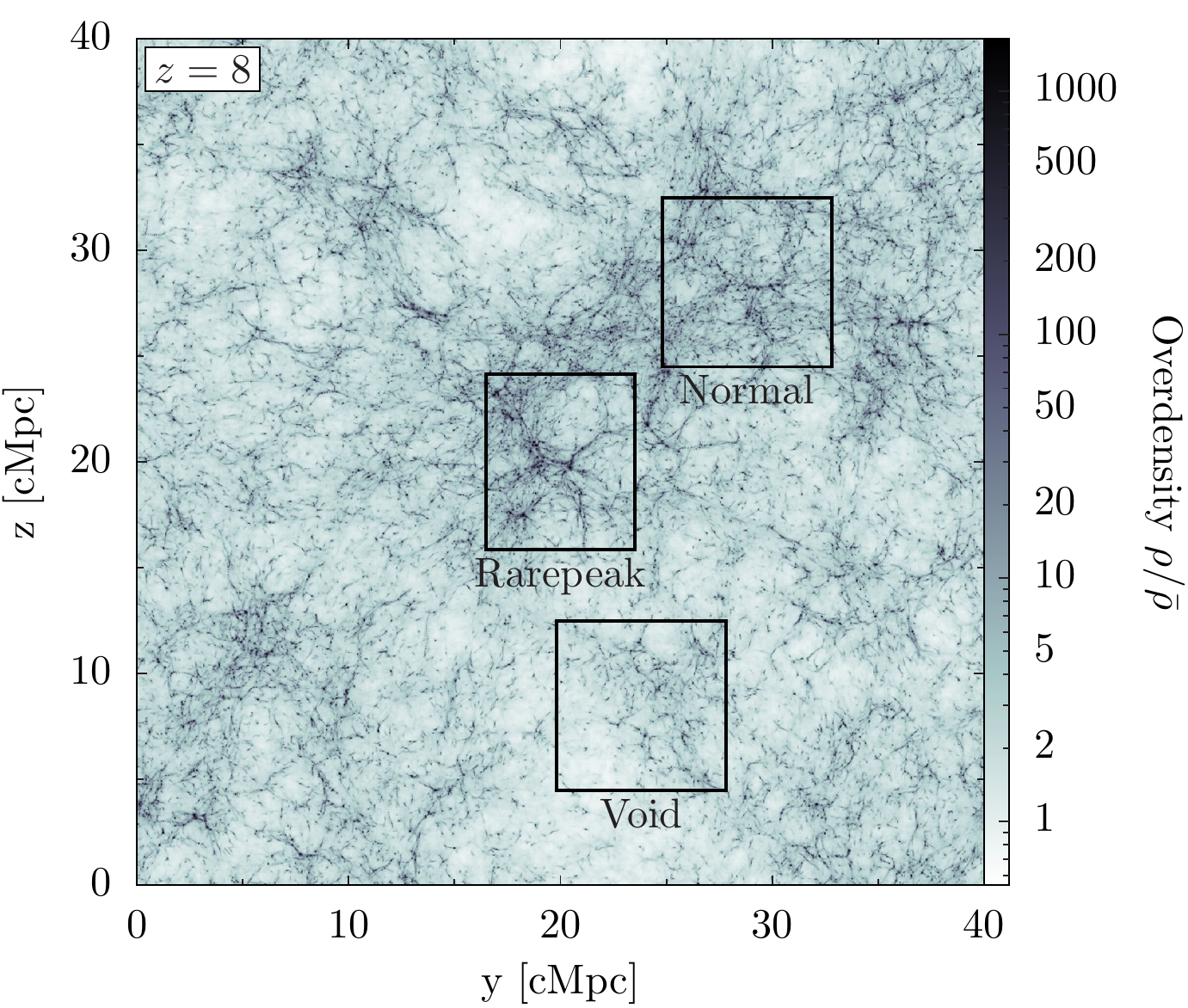}
    \caption{Mass-weighted density projection of the (40 comoving Mpc)$^3$ exploratory dark matter simulations at redshift $z=8$ used in the \renaissance suite. The survey volumes of the Rarepeak, Normal and Void regions are outlined. The Rarepeak region is centered on the most massive halo at $z = 6$. Due to projection effects, the Normal region appears as dense as the Rarepeak region. However, the average overdensity of the Normal region is only 9\% higher than the mean matter density. Visualisation originally published in \cite{Xu_2016}.}
    \label{fig:Renaissance_suite}
\end{figure}
% \vspace{1cm}
\section{Methodology}
\label{method}
\subsection{The Renaissance Simulations}
\noindent The \renaissance simulations \citep{renaissance, chen_renaissance, OShea_2015, Smith_2018, Wise_2019} were run using the massively parallel adaptive mesh refinement \enzo code \citep{Enzo_2014, Enzo_2019}. We briefly describe the methods used here, but refer the interested reader to the previous papers for a more complete discussion. The \renaissance simulation suite is composed of three zoom-in regions (see Figure \ref{fig:Renaissance_suite}) extracted from a parent {\color{black}comoving} volume of 40 Mpc on a side. The three separate zoom-in regions were named the Rarepeak (RP) region, the Normal region and the Void region. The zoom-in volumes ranged from 200 to 430 comoving Mpc$^3$. \\
\indent The RP region is centred on a $3 \times 10^{10}$ \msolar halo at $z = 6$ with an enclosing 
{\color{black}comoving} volume of $(3.8 \times 5.4 \times 6.6)$ Mpc$^3$. The Normal and Void volumes have comoving volumes of $(6.0 \times 6.0 \times 6.125)$ Mpc$^3$. All three regions have projected areas comparable to the NIRCam field of view {\color{black}with the RP, Normal, and Void regions respectively subtending ($1.2^\prime \times 1.8^\prime \times 2.2^\prime)$ at $z=15$, ($2.1^\prime \times 2.1^\prime \times 2.1^\prime)$ at $z=11.6$, ($2.2^\prime \times 2.2^\prime \times 2.3^\prime)$ at $z=8$}. The \renaissance suite uses the cosmological parameters from the 7-year WMAP \LCDM+SZ+LENS best fit \citep{Komatsu_2011} with $\Omega_{\rm M} = 0.266, \Omega_{\Lambda} = 0.734, \Omega_{\rm b} = 0.0449, h = 0.71, \sigma_8 = 0.81$ and $n = 0.963$. The (dark matter) particle mass resolution of the \renaissance suite is $2.9 \times 10^4$ \msolar and the maximum spatial resolution 
afforded by the adaptive mesh is 19 comoving pc. This allows the \renaissance suite to resolve most of the minihaloes in which
the first stars are expected to form \citep[e.g.][]{Machacek_2001, Kulkarni_2021, Chiaki_2023}. \\
\indent In particular, \renaissance employs a model for metal-free (Population III) and metal-poor (Population II) star formation \citep{Wise09, Wise_2012b} allowing for the stochastic sampling of the formation of the first stars and galaxies{\color{black}, inspired by star formation simulations and observations rather than their galactic counterparts because they resolve individual star-forming clouds.  Under the assumption that metal-enriched star formation is insensitive to redshift, \citet{Wise_2012b} used an efficiency of 7 per cent of the cold gas within a dense cloud being converted into Population II star particles, similar to what was found in present-day star formation simulations \citep{Krumholz05}, which were themselves consistent with observations \citep{Tan06} at the time of the method's formulation.}  They calibrated the feedback models against local dwarf galaxy properties, in particular the metallicity distribution function, ensuring that they avoided the ``overcooling problem'' that overestimates the stellar mass and thus metallicities.  The resulting metal enrichment driven from the collapse of the first stars results in the emergence of the second generation of stars which ultimately lead to the birth of the first massive galaxies - the galaxies which JWST is now observing. \\
\indent The computational complexity of the \renaissance suite means that evolving these simulations to the present day is completely intractable. As such the RP simulation was evolved to $z = 15$, the Normal simulation evolved to $z = 11.6$ and the Void region to $z = 8$. As some of the JADES and CEERS results are at somewhat lower redshifts (compared to the RP and Normal runs) we extrapolate our results to the JADES and CEERS spectroscopic redshifts in some cases. \\
\indent As discussed in the Introduction, the comparison study undertaken by K23 uses the simulation datasets from EAGLE, Illustris, TNG100, RomulusC, {\sc Obelisk} and Simba. In Table \ref{simulations} we compare the 
simulation datasets used in K23 versus that used here (i.e. against \texttt{Renaissance}). The simulations used in K23 do not have sufficient resolution to probe the formation of the first stars and can resolve, at best, the formation of the first 
atomic cooling haloes. The \renaissance suite allows us to probe the assembly processes involved in forming the haloes that appear in the simulations used in K23 as well as the building blocks of the galaxies now being observed with JWST.
\begin{table}[]
    \centering
    \caption{Simulation domain sizes and resolution}
    \label{simulations}
    \begin{tabular}{|c|c|c|c|}
        \hline
        Simulation & Box size [cMpc]& $M_{\rm DM}$ [M{$_\odot$}] & $\Delta x_{\rm DM, *}$ [pc]\\
        \hline
        Simba  & 147.7 &$9.7\times 10^7$& 500\\ 
         EAGLE & 100 &$9.7 \times 10^{6}$  & 2660\\
         TNG100  & 110.7 &$7.5\times 10^6$ & 740 \\
         Illustris & 106.5 &$6.3\times 10^6$  & 710\\  
         {\sc Obelisk}  & 142.0 &$1.2\times 10^6$ & 540\\
         RomulusC  & 50 & $3.4 \times 10^5$  & 250\\
         Renaissance  &  40 & $2.4\times 10^4$  & 19\\
         \hline
    \end{tabular}
    \parbox[t]{0.95\columnwidth}{\vspace{1ex}\textbf{Notes:} 
    The first column gives the simulation suite name, the second column the comoving box size length of each simulation used in K23, the third column gives the dark matter particle resolution, $M_{\rm DM}$,  and finally the fourth column gives the spatial resolution, The spatial resolution was based on the gravitational softening lengths for the SPH simulations and for the AMR simulations it is based on the cell length. For the Renaissance suite we give the parent box size (40 cMpc) but note that the results we show here are for the zoom in regions which have box lengths of approximately 6 cMpc.
    }
\end{table}

\subsection{Extrapolating the Stellar Mass of the Simulated Galaxies based on the Star Formation History}
\label{Extrapolating}
\noindent To properly compare the simulated galaxy conditions against observations, we need to have simulated values at the same redshift as the JADES and CEERS measurements. However, as discussed, both the Normal and RP regions of the \renaissance simulations only reach a redshift of $z = 11.6$ and $z = 15$ respectively. Therefore, the Normal region does not reach sufficiently low redshifts so as to be directly comparable against two of the JADES and CEERS galaxies, while the RP region cannot be directly compared, in terms of redshift, against any of the JADES and CEERS galaxies. To rectify this, we extrapolate the stellar masses of the most massive galaxy in both the Normal and RP region based on their respective {\color{black}{specific Star Formation Rates (sSFR)}} forward in time to connect with the observational redshifts.\par 
The definition of the sSFR is
\begin{equation}
    \Psi_{\rm S} \equiv \frac{\Psi}{M_{*}}
\end{equation}
where $\Psi$ is the SFR and $\rm{M_*}$ is the stellar mass. {\color{black} We use three values for $\Psi_{\rm S}$ in our extrapolation method: A maximum value of $10^{-7}$ yr$^{-1}$, a nominal value of $10^{-8}$ yr$^{-1}$, and a minimum value of $10^{-9} \; \textrm{yr}^{-1}$. These values were chosen from the range of sSFR values found in \renaissance (see also Figure \ref{fig:Mstar_vs_SFR_plot}). The extrapolated mass is dictated by the differential equation 
\begin{equation}
    \Psi \equiv \frac{dM_*}{dt} = \Psi_{\rm S}  M_*,
\end{equation}}
where the solution of this equation is
\begin{equation} \label{Eqn:M*}
    M_*(t) = M_0 e^{\Psi_{\rm S} \left(t - t_0\right)}.
\end{equation}
Here, $M_0$ is the final simulated stellar mass of the halo and $t_0$ is the final simulated time. Using Equation (\ref{Eqn:M*}) we can then predict the stellar mass of the galaxy past the 
final simulation time. We will discuss the impact of this extrapolation method in more detail next.

\section{Results}
\label{results}

\noindent We begin here by comparing the most massive galaxy in each of the RP, Normal and Void regions against the JADES and CEERS results. We then follow this by comparing the global galaxy assembly history of all of the galaxies in the \renaissance suite against the JADES and CEERS results. 
\begin{figure*}[]
    \centering
    \includegraphics[width=0.9\textwidth]{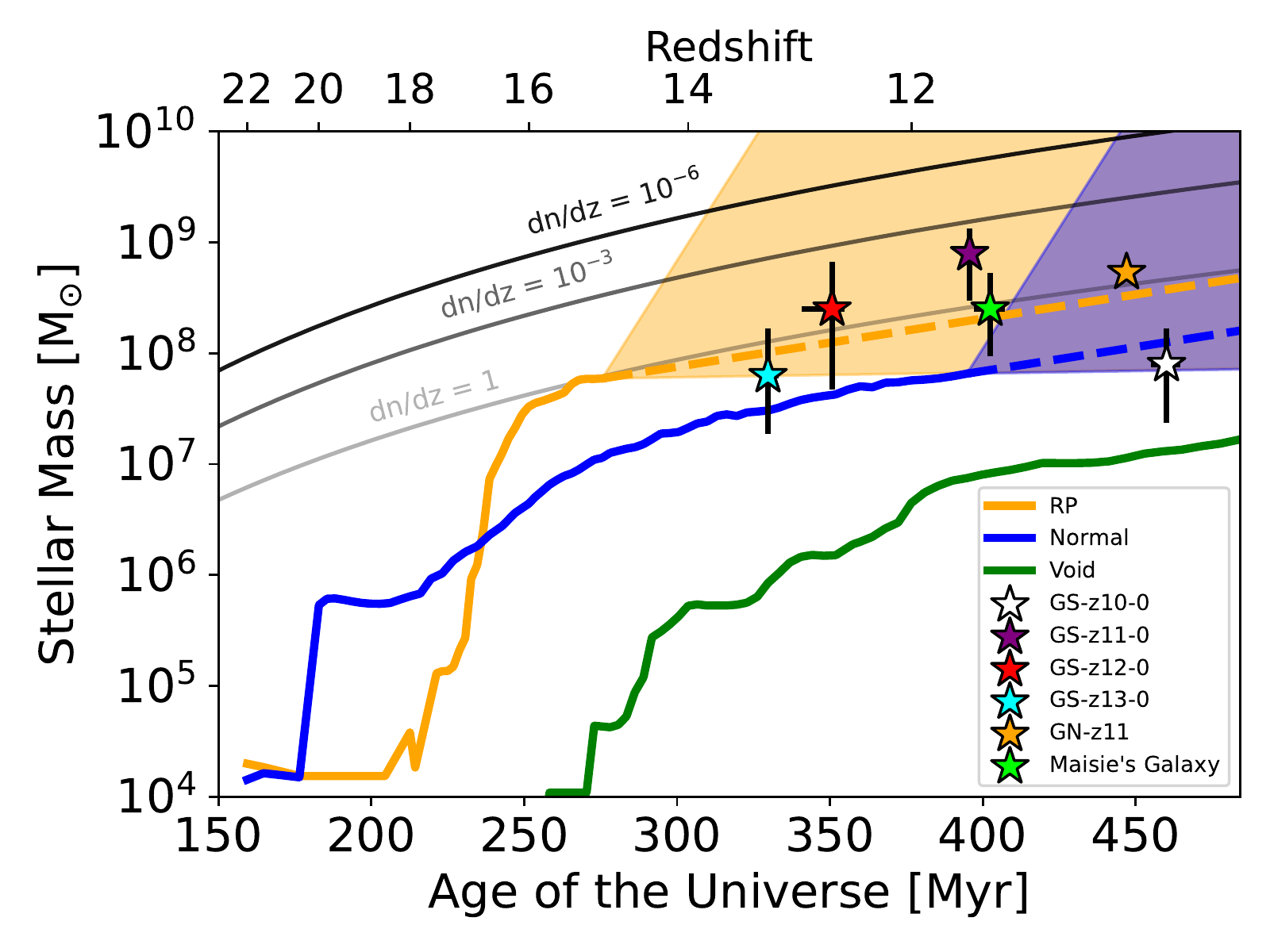}
    \caption{The most massive galaxy in each of the Rarepeak (RP, orange), Normal (blue) and Void (green) regions are shown. The six star shaped symbols identify the JADES and CEERS galaxies giving both their stellar masses and the redshift at which they were spectroscopically identified. The most massive galaxies in both the RP and the Normal regions are in excellent agreement with the JADES and CEERS observations. {\color{black}The shaded region denotes the extrapolation of the stellar masses based on the SFR history of the respective galaxy where the region is bounded from  above by a specific star formation rate of $10^{-7}\; \textrm{yr}^{-1}$ and from below by a sSFR of $10^{-9}\; \textrm{yr}^{-1}$. The dashed line represents an extrapolated mass based on a sSFR of $10^{-8}\; \textrm{yr}^{-1}$}. Finally, we plot lines {\color{black}labelled by $dn/dz$} the expectation value for finding a galaxy of a given stellar mass at a given redshift in a field of view comparable to NIRCam. For this we assume a gas mass corresponding to the baryon fraction and a star formation efficiency (SFE) of 0.1 using the Sheth-Tormen \citep{ShethTormen_1999} halo mass function as the underlying framework. All of the JADES, CEERS and Renaissance haloes are consistent with $\Lambda$CDM predictions of finding at least one halo with these stellar masses in this volume. See text for further details on the consistency with respect to $\Lambda$CDM predictions.}
    \label{fig:Stellar_mass_plot}
\end{figure*}
\subsection{Comparing the Most Massive Galaxies in Renaissance with the JADES and CEERS measurements}
\label{most_massive}
\noindent In Figure \ref{fig:Stellar_mass_plot} we plot the stellar masses of the most massive galaxy in each region in \renaissance against the stellar masses of each of the JADES and CEERS survey galaxies. {\color{black}As discussed in \S \ref{Extrapolating}, we extrapolate the stellar masses to the observed redshifts for the comparison.}\\
\indent From Figure \ref{fig:Stellar_mass_plot} we can immediately see that the most massive galaxy in the Normal region (blue line), which evolves to a $z = 11.6$, has a stellar mass greater than GS-z10-0 and is consistent with or within a factor of a few compared to the stellar mass of the remaining five galaxies. The RP region, which does not overlap in redshift with any of the JADES and CEERS galaxies, is consistent in terms of stellar mass with each of the observed galaxies {\color{black}once extrapolation (using the sSFR) is considered}. The upper bounds of both shaded regions is based on a constant sSFR value of $10^{-7} \textrm{ yr}^{-1}$, the lower bounds being $10^{-9} \textrm{ yr}^{-1}$, and the dashed lines show the extrapolated stellar masses based on an nominal sSFR value of $10^{-8}$ yr$^{-1}$. The dashed line belonging to the RP region is {\color{black}overall} consistent with each of the observed galaxies{\color{black}, with GS-z11-0 and GN-z11 lying within the bounds of the extrapolations}. The Void region shows systematically lower stellar masses and does not achieve the same stellar masses as those found in JADES and CEERS. \\
\indent Large-scale overdensities are directly related to the overabundances of galaxies through the halo mass function and the halo mass -- stellar mass relation, shown by \citet{Xu_2016} for the \renaissance suite in particular.  The observed stellar mass estimates suggest that the CEERS and JADES fields are not likely underdense regions. However, additional work would be needed to statistically conclude that the JADES and CEERS were observed in overdense regions.\par
\begin{figure}[!t]
    \centering
    \includegraphics[scale = 0.53]{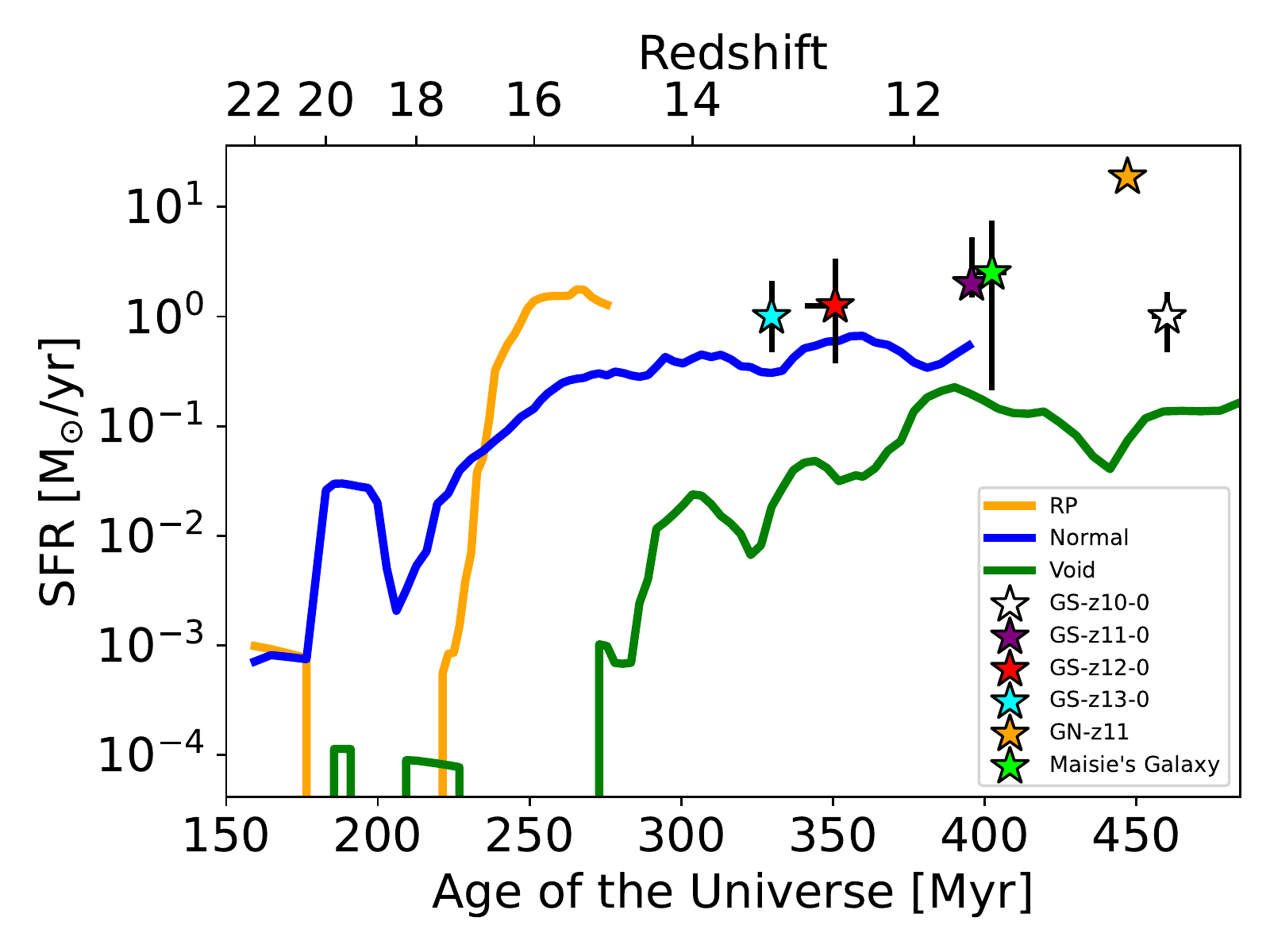}
    \caption{The star formation rates (SFRs) of the most massive galaxy in each of the RP (orange), Normal (blue) and Void (green) regions. The SFR of the JADES and CEERS galaxies are also plotted as star symbols with associated error bars \citep{robertson, Maisie, bunker2023jades}. All of the modelled galaxies show an initial burst of star formation at the early stages of evolution with a leveling off as each halo continues to evolve. In the cases where the observed galaxies overlap with the modelled galaxies, the SFRs found are consistent with 
    each other within an order of magnitude. Both the Normal and RP regions show SFRs, for their highest mass galaxies of between 0.5 and 1 \msolar yr$^{-1}$. The most massive galaxy in the Void region shows a lower SFR as expected \citep[e.g.][]{Xu_2016}.}
    \label{fig:SFR_plot}
\end{figure}

 
\indent To provide additional context and to quantify the rarity of haloes at these masses and redshifts, we plot lines in Figure \ref{fig:Stellar_mass_plot} to represent the number of haloes of a certain (stellar) mass we expect JWST to see at a specific redshift. These lines are based on the baryon fraction obtained from WMAP7 and a star formation efficiency (SFE) of $f_* \equiv M_*/[(\Omega_{\rm b}/\Omega_{\rm M}) M_{\rm halo}] = 0.1$,{\color{black}with the baryonic fraction being $\Omega_b/\Omega_M \sim 0.17$. While this value for the SFE may seem high \citep[see e.g.][]{SFE-Tacchella} for this redshift regime, \renaissance shows an increasing SFE value dependent on the halo mass (see Figure \ref{fig:Halo_vs_Stellar_mass_plot}), for this reason a value of 0.1 seems appropriate for the extrapolation}. The value of $dn(M_*, z)/dz$ represents the number of galaxies of stellar mass $M_*$ or greater we expect to see at redshift $z$ in the NIRCam field of view. For example $\rm{dn/dz} = 10^{-3}$ tells us that we expect to see one galaxy with stellar mass $M_*$ or greater at redshift $z$ in every one thousand frames with the angular size of NIRCam.\\
\indent All of the galaxies are consistent with the $\rm{dn/dz} = 1$ line except the GS-z11-0 galaxy whose error bars are marginally above the $\rm{dn/dz} = 1$ line. This means that we should expect to observe at least one galaxy of that stellar mass in the field of view of NIRCam. The lines at $\rm{dn/dz} = 10^{-3}$ and $10^{-6}$ represent a mass scale that would be increasingly unlikely to observe at that redshift. The calculations used in the creation of these values is described in Appendix \ref{Expected_halo}.\\
\indent {\color{black}We can observe that the Normal region does not contain a galaxy that exceeds the $dn/dz = 1$ line. This can, however, be explained through cosmic variance \citep{cosmic_variance_chen, Cosmic_variance_Bhowmick_2020} which would provide an uncertainty in the expectation values of the stellar mass of $\sim\pm$100\% \citep[e.g.][]{cosmic_variance_calc}. It may also be explained by a variance in the halo mass function used at high redshifts \citep{Cosmic_variance_yung2023ultrahighredshift}, which would provide an additional uncertainty up to an order of magnitude in the abundances of halos {\color{black} at fixed stellar mass}. Combining these two uncertainties can easily move the $dn/dz = 1$ lines up (or down) by a factor of a few.}\\
\indent {\color{black}In summary, the JWST observed galaxies are overall consistent with both the Normal and RP regions once the extrapolation methods and cosmic variance are taken into account. Both GN-z11 and GS-z11-0 are unusual galaxies and detailed hydrodynamic simulations are clearly struggling to match their luminosities and stellar masses perfectly. It may be that these galaxies contain star formation processes not currently correctly modelled by our subgrid recipes - {\color{black}by this we mean that these JWST galaxies may contain stars born from an extremely top heavy IMF including a population of extremely massive stars \citep[e.g.][]{Charbonnel_2023, Nagele_2023} or unusually strong bursts of star formation \citep[e.g.][]{Kobayashi_2023}}. In addition to this, GN-z11 shows signatures of AGN activity \citep{maoilino-agn, GNz11-AGN-scholtz} which is not currently modelled by \renaissance for example {\color{black}and the processes responsible for seeding this massive black hole may well be a factor in explaining its large luminosity}. Finally, we make no attempt here to analyse absolute cosmic abundances of high mass galaxies. The galaxies found thus far by JWST at these redshifts are at the absolute limit of the Renaissance simulations and we are not in a position to quantify such abundances.} {\color{black}This is due to the small volume of the Renaissance simulations and thus the limitation in quantifying the statistics of high mass haloes in this epoch of the Universe.} \\ 
\indent Figure \ref{fig:SFR_plot} shows the SFR history of the most massive galaxy in the RP, Normal and Void regions - the same as Figure \ref{fig:Stellar_mass_plot}. Here and thereafter we consider the average SFR over the previous 20 Myr, i.e. the stellar mass of star particles younger than 20 Myr divided by 20 Myr.
Comparing the observed (coloured symbols) and simulated galaxies (solid coloured lines), it is clear that \texttt{Renaissance} is able to reproduce galaxies with star formation rates generally consistent with the galaxies observed in JADES and CEERS. This is achieved due to the excellent mass and spatial resolution afforded by \renaissance which gives \renaissance the unique capability to follow the formation of the galaxy stellar population(s) from approximately $10^4$ \msolar up to the final simulated stellar masses of a few times $10^7$ \msolarc. What we see is that the star formation rates start several orders of magnitude below that observed but in very inefficient, star forming, haloes. As the star formation efficiency increases (see also Figure \ref{fig:Mstar_vs_SFR_plot}) the simulated values quickly {\color{black}level off} to the observed values ultimately reaching values of approximately 1 \msolar yr$^{-1}$.\\
\indent It is the higher mass resolution of \texttt{Renaissance}, afforded through the use of adaptive grids in a zoom-in setup, that allows us to follow the build up of the galaxy from initially very small stellar masses ($\sim $ $10^4$ M$_\odot$) up to a final stellar mass of more than $>10^7$ M$_\odot$. Additionally, the spatial resolution of 19 comoving pc allows for the internal structure and dynamics of these early galaxies to be well resolved.\\
\indent We observe an initial burst of star formation in the \renaissance galaxies plotted in Figure \ref{fig:SFR_plot} with a leveling off in the SFR as the halo continues to evolve. This {\color{black}can be explained via the following mechanisms}. In the initial stages of star formation in the haloes, the mass resolution of \renaissance is sufficient to capture the early assembly history. Once the halo mass exceeds the atomic cooling threshold and gas can cool via atomic line emission, the gas begins to collapse and form Population II stars more readily and the star formation efficiency increases rapidly. The feedback from the Population II star formation heats up the surrounding environment thus regulating further star formation. This results in an equilibrium in the SFR of the halo and is represented by the leveling off seen in Figure \ref{fig:SFR_plot}.

\subsection{Comparing all Galaxies in Renaissance with JADES Measurements and Simulation Results} 

\begin{figure}[!t]
    \centering
    \includegraphics[scale = 0.365]{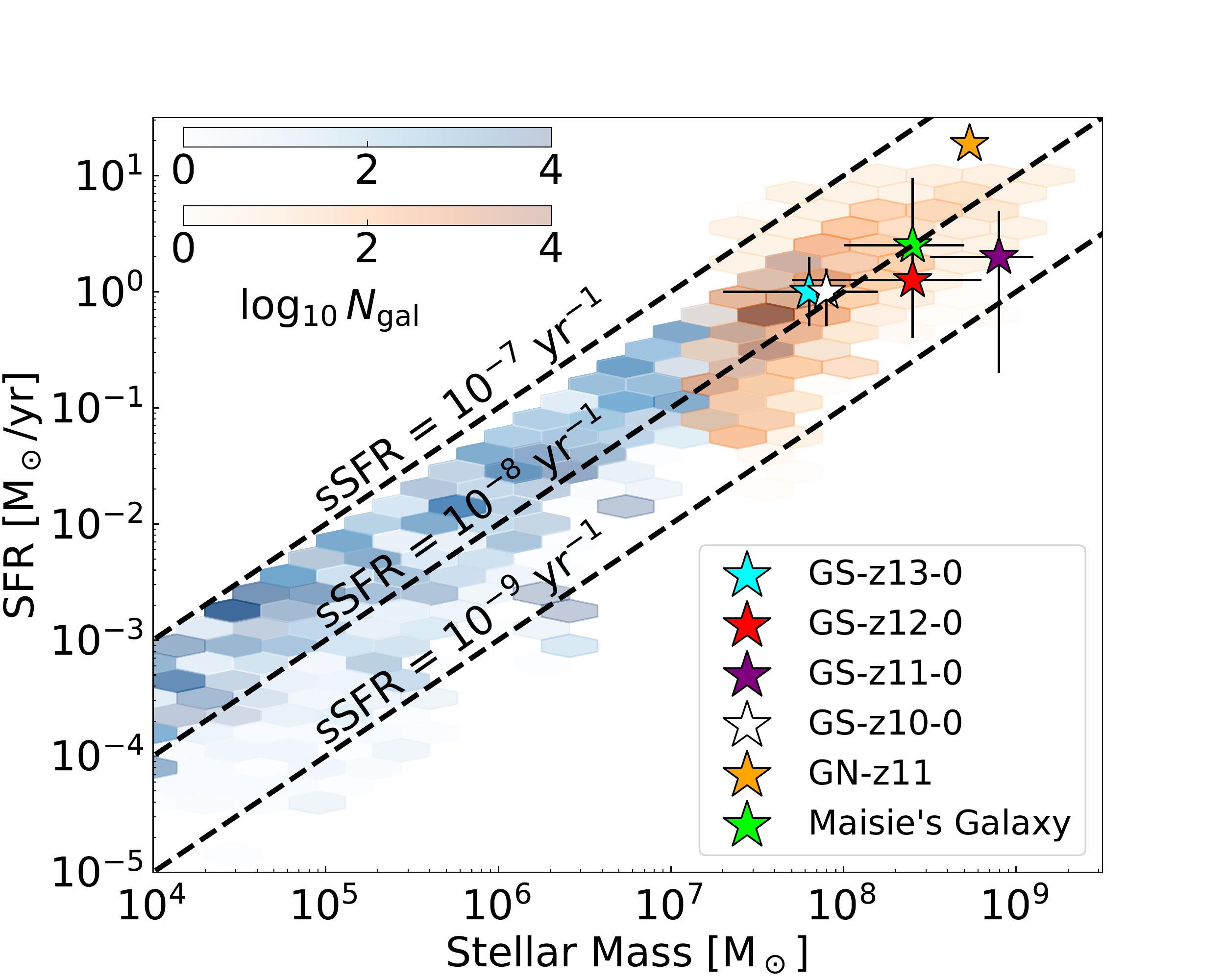}
    \caption{The star formation rates of the galaxies in the Renaissance simulations, which occupy lower masses in the plot and shown in blue hexbins, and the galaxies in the K23 simulations, which exist at higher masses and shown in orange hexbins, as a function of stellar mass. {\color{black}The bin saturation depicts the number of galaxies in a log-scale. The galaxies from the \renaissance simulations (blue hexbins) exist at redshifts greater than 15, 11.6, and 8 for the RP, Normal, and Void region, respectively. The K23 simulations are in the range of $z = 10-14$ for four full-box and two zoom-in simulations. We consider all outputs from the simulations as a stack over all snapshots. } Lines of constant sSFR are plotted that are representative of the majority of the galaxies in the simulations. We plot the JADES and CEERS observations as stars, consistent with the simulation data.}
    \label{fig:Mstar_vs_SFR_plot}
\end{figure}

\noindent We now expand our analysis to a larger set of high-redshift simulated galaxies to determine how the observed galaxies compare.  Figure \ref{fig:Mstar_vs_SFR_plot} shows the SFR of each galaxy as a function of the stellar mass of each galaxy in \renaissance along with the galaxies from the simulations included in K23. By including the data from our simulations and the K23 simulations, we see how both lower and higher mass galaxies compare to the JADES and CEERS data that is plotted on Figure \ref{fig:Mstar_vs_SFR_plot} as coloured stars. The data on GS-z13-0, GS-z12-0, GS-z11-0 and GS-z10-0 are taken from \cite{robertson}, GN-z11 is taken from \citet{bunker2023jades} while the data on Maisie's Galaxy is taken from \cite{Maisie}. In this comparison, we consider all galaxies in the RP, Normal, and Void regions of the \renaissance simulations that extend to very low stellar masses, 
shown starting from $10^4$ \msolarc{} (blue hexbins), while the K23 simulations (orange hexbins) 
provide insight into the higher mass regime. % As shown by Figure \ref{fig:Mstar_vs_SFR_plot} 
%the \renaissance and K23 simulations' galaxies align well with the JADES and CEERS data. 
{\color{black}As shown by Figure \ref{fig:Mstar_vs_SFR_plot}, the trends found in the \renaissance simulations compliment the K23 simulations at higher masses that overlap the JADES and CEERS observations, showing the cohesion between the simulations and observed data.}

\par
We show lines of constant sSFRs that are representative of the majority of the data in Figure \ref{fig:Mstar_vs_SFR_plot}. The figure shows that most galaxies at high-redshift quickly assemble, lying in the range $10^{-7} - 10^{-9}\; \mathrm{yr}^{-1}$ corresponding to $e$-folding times of 10 Myr to 1 Gyr.  While it is difficult to see any definite trends in the heatmap because of the differing simulations and zoom-in regions, this figure demonstrates that early galaxy formation progresses at a fairly rapid rate with the lower mass galaxies having higher sSFR values. 
The plot shows that the \renaissance simulation follows closely with the trend shown in the K23 simulations. For both sets of data, we can see that the \jwadd{distribution of the} SFRs increases as stellar mass increases with a constant slope. The \renaissance data connects smoothly to the K23 and JADES data, showing how the galaxies in the \renaissance simulations could evolve over 200 Myr to become the JADES and CEERS galaxies. %The figure clearly demonstrates that the \renaissance and K23 simulations are consistent with the observed sSFRs of $\sim 10^{-8} \; \mathrm{yr}^{-1}$ and that these galaxies are not out of the ordinary.
{\color{black}The consistent progression of SFR in Figure \ref{fig:Mstar_vs_SFR_plot} from the low-mass regime (blue hexbins) to higher masses (orange hexbins) indicates that these $z>10$ galaxies detected by JADES and CEERS are the most massive examples within a much larger and still undetected population of young and low-mass galaxies.}

\subsection{Simulated Stellar Mass -- Halo Mass Relation}

\begin{figure}[!t]
    \centering
    \includegraphics[scale = 0.365]{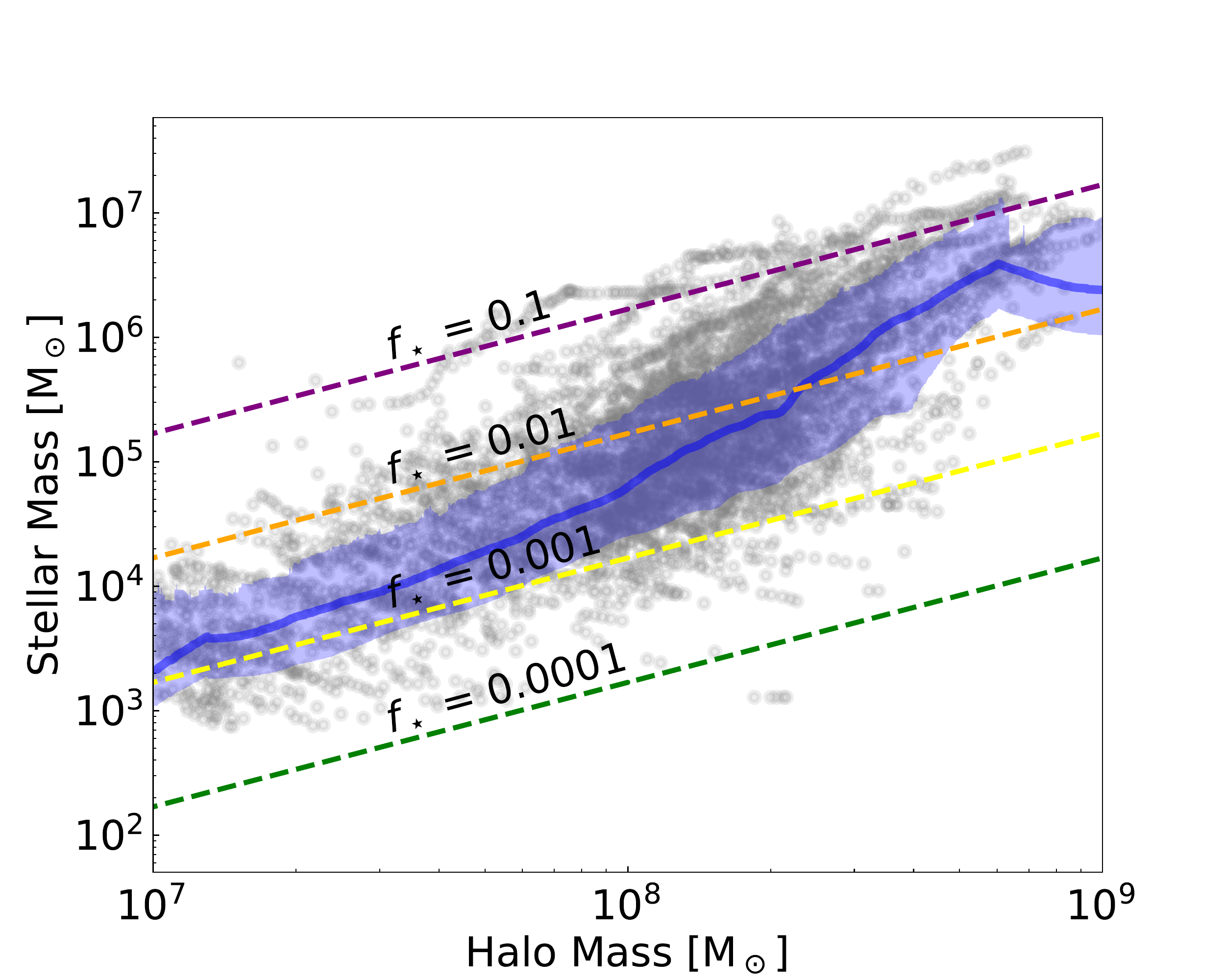}
    \caption{The stellar mass of each galaxy from the Renaissance simulation plotted against the halo mass of each galaxy over time. The green, yellow, orange, and purple dashed lines show the stellar mass given a constant SFE of 0.01, 0.1, 1, and 10 per cent, respectively.  The blue solid line shows the median stellar mass while the shaded blue region shows the standard deviation, each of them as a function of the halo mass.}
    \label{fig:Halo_vs_Stellar_mass_plot}
\end{figure}

\jwadd{The growth of low-mass galaxies is ultimately controlled by the host dark matter halo growth, thus we turn our attention to the stellar mass - halo mass relationship of the galaxies in the Renaissance Simulations to shed light on the origins of the galaxies observed in CEERS and JADES.} Figure \ref{fig:Halo_vs_Stellar_mass_plot} shows the stellar mass of each galaxy in the \renaissance simulation against its host halo mass from all zoom-in regions {\color{black}which shows the linear relationship between the halo mass and the stellar mass in the \renaissance galaxies until reaching the atomic cooling limit}.  We plot data from all outputs in the simulation suite.  \jwadd{We can combine these datasets because galaxy scaling relations do not vary greatly, though with a large galaxy-to-galaxy scatter, with redshift at these early times \citep[see Figure 15 and Table 2 in][]{Xu_2016}.} Here we focus on the trends between stellar mass and halo mass in this low-mass regime.  From the raw data alone, there is a large scatter at all halo masses, culminating from highly variable SFRs that arise from periodic gas expulsions from stellar feedback.

To better show any trends, we show the median value (solid blue line) as a function of the halo mass for all galaxies. We depict the standard deviation as the shaded blue region, which remains fairly constant in log-space throughout this mass range.  We also show lines of constant star formation efficiency (SFE) \jwadd{for convenience}. As demonstrated in Figure \ref{fig:SFR_plot}, galaxy growth does not proceed at a constant SFR and thus SFE, especially at these low masses when star formation can be bursty.  Each galaxy makes its unique path through this parameter space.  When a population of bursty galaxies are considered, these variable tracks transform into a relationship shown in the figure. Throughout most of the plot, the slope of the median line is steeper than the slopes of the constant SFE lines, meaning that the galaxies are, on average, forming stars more efficiently as they grow.

This trend clearly shows an increase in the $M_* - M_{\rm halo}$ slope around the atomic cooling limit at $\sim 10^8 \; {\rm M}_\odot$.  Gas cooling and thus star formation within these nascent galaxies are inefficient below this limit with a median SFE $f_* \sim 2 \times 10^{-3}$.  Once the gas can cool through atomic processes, star formation becomes more efficient, depicted by the increase in the slope and associated SFE.  By the time haloes reach $10^9\; {\rm M}_\odot$, the median SFEs are a few per cent, similar to the more massive galaxies probed by the simulations highlighted in K23. {\color{black}With this information, we are able to justify our use for the SFE value of SFE = 0.1 that we use for the mass extrapolation discussed in \S\ref{Extrapolating}}.

We note that the turnover to a negative slope is an artefact of limited galaxy sample at these highest halo masses in the \renaissance simulations.  In principle, the stellar mass should continue to grow as the halo grows.  These most massive haloes have lower stellar masses than some of slightly less massive haloes.  This is not unexpected because of the large scatter in the stellar mass - halo mass relation, caused by the stochastic nature of early galaxy formation.

\section{Discussion and Conclusions}
\noindent The goal of this study is to investigate whether or not the initial findings of JWST, via the JADES and CEERS surveys, are consistent with state-of-the-art high resolution simulations. Additionally, the high (mass) resolution simulations allow us to study in detail the assembly history of these galaxies and allow us to connect the 
modelled galaxies with those observed in JADES and CEERS.
We find that, using the high spatial and mass resolution \renaissance simulations, that excellent agreement exists between observations and simulations. {\color{black}There is a limitation in the fact that we are unable to compare the simulations and measurements at similar redshifts. To address this, an extrapolation method was applied to the end masses of the most massive haloes in the RP and Normal regions. We justify this caveat, by extracting directly the range of sSFR values found across all halo mass ranges in \renaissance (see Figure \ref{fig:Mstar_vs_SFR_plot}). These sSFR values were then used to put lower and upper bounds on the mass of the most massive haloes in both the Normal and RP regions}.\\
\indent {\color{black}We also address the fact that both the GN-z11 and GS-z11-0 galaxies are above the (extrapolated) mass of the most massive galaxy in the RP region (when a sSFR of $10^{-8}$ yr$^{-1}$ is used). GN-z11 was recently found to have an anomalously high abundance of N/O \citep{GN-z11-NO-abundance} and likely hosts an active AGN \citep{Maiolino_2023, GNz11-AGN-scholtz}, a characteristic that will be missed by \texttt{Renaissance}. These outliers may also be explained by an uncertainty associated with cosmic variance, as discussed in \S \ref{most_massive} or an uncertainty associated with the higher ends of the redshift ranges \citep{Cosmic_variance_yung2023ultrahighredshift}}.\par
Our results are consistent with a similar study by \cite{Keller_2023} who compared a range of somewhat larger scale simulations against the initial JADES and CEERS results. {\color{black}The combination of the larger scale study by K23 with the higher resolution 
simulations here, which can accurately track the initial assembly history of the galaxies subsequently found in the K23 simulations, provides compelling evidence for excellent agreement between the JADES and CEERS results and the $\Lambda$CDM model. It is the combination of the two regimes that is crucial in this regard.}
Our overall findings can be broken down as:
\begin{itemize}
    \item The most massive haloes in \renaissance have comparable stellar masses to the JADES and CEERS galaxies after extrapolating the masses. {\color{black}The Renaissance simulation suite does not evolve in time to the same epoch as all of the observations and so we must extrapolate the (specific) star formation rate forward in time for some galaxies - but nonetheless the agreement is overall remarkable.} Comparing with the theoretical expectation of galaxies within a field of view identical to NIRCam, we find that the $z > 10$ galaxies detected in JADES and CEERS are consistent with that expected from a $\Lambda$CDM cosmology.
   
    \item The star formation rates for the most massive galaxies in \renaissance are {\color{black}overall} consistent with the JADES and CEERS measurements. The star formation histories show specific star formation rates, as a function of stellar mass, generally consistent with these latest $z > 10$ JWST observations. {\color{black}Finally, it is also possible that the observed stellar masses may be affected by bursts of star formation close to the epoch of observation which may effect the derived results \citep{narayanan2023outshining}}.

    \item  The mass resolution of \renaissance allows us to capture the rapid assembly of galaxies in the early Universe.  After inefficiently forming stars below the atomic cooling threshold at a median SFE of $2 \times 10^{-3}$, star formation becomes more rigorous yet feedback-regulated, reaching levels of a few per cent at galaxy masses  similar to the JADES and CEERS measurements.
    
    \item We conclude that both lower and finer resolution simulations agree that the JADES and CEERS measurements and are not in tension with current galaxy formation models.
\end{itemize}

% \sout{the stellar masses of the most massive galaxies in the Rarepeak and Normal regions need to be extrapolated so that they can be compared to the observational measurements. After extrapolation, we find that the galaxies need to maintain an sSFR of at least $10^{-8}$ yr$^{-1}$ in order to have a stellar mass comparable to the galaxies in the large scale simulations and the JADES and CEERS observations.} 

\noindent JWST has for the first time enabled a detailed view of the early Universe. Initial findings of massive early galaxies has surprised many with some discussion in the literature that the JWST results maybe in conflict with $\Lambda$CDM \citep[][]{Haslbauer_2022, boylankolchin2023stress}. However, what we find is that in the context of a $\Lambda$CDM Universe there is no tension between theory and observation at the very highest redshifts that we can currently probe. 

As more measurements are made with JWST and future record breaking observatories, it comes with more opportunity to utilise high-resolution simulations like \texttt{Renaissance} and further stress test the $\Lambda$CDM model.
\label{conlusions}
\section*{Acknowledgements}
\noindent We thank the referees for providing constructive feedback and also Peter Coles for useful discussions during the course of this work. We also thank Ben Keller for providing the data points used in Figure \ref{fig:Mstar_vs_SFR_plot}.
JM acknowledges the support from the John \& Pat Hume Doctoral Awards Scholarship (Hume 2021-22).  JR acknowledges support from the Royal Society and Science Foundation Ireland under grant number 
 URF\textbackslash R1\textbackslash 191132. JR also acknowledges support from the Irish Research Council Laureate programme under grant number IRCLA/2022/1165. JHW acknowledges support by NSF grants OAC-1835213 and AST-2108020 and NASA grants 80NSSC20K0520 and 80NSSC21K1053.
\bibliographystyle{mnras}
\bibliography{references}
\appendix
\section{Appendix A: Calculating the number of haloes expected to be seen by JWST}
\label{Expected_halo}
\noindent In Figure \ref{fig:Stellar_mass_plot} we present predictions for the number of haloes of a certain stellar mass, $M_*$ we expect to see in JWST's FoV, $\Omega$, at a certain redshift, $z$. To do this we use the \texttt{hmf} module \citep{2013A&C.....3...23M} to supply a halo mass function which we then use in our calculation. The halo mass function provides the number of haloes with a mass between $M$ and $M + dm$ at $z$ per comoving volume. The number of haloes we expect to see per comoving Mpc in JWST's field of view, with a mass greater than $M_0$, is calculated as follows:
\begin{equation}
    \frac{dn}{dr_c} = A\left( z \right) \int_{M_0}^\infty \frac{dn}{dm}\left( M, z\right)dm
\end{equation}
$A(z)$, measured in Mpc$^{2}$, is the comoving area at redshift $z$ and is calculated from
\begin{equation}
    A(z) = \Omega r_c^2(z)
\end{equation}
where $r_c(z)$ is the comoving distance as a function of redshift. We can then convert it to the amount of haloes NIRCam should observe per redshift with 
\begin{equation}
    \frac{dn}{dz} = \frac{dn}{dr_c}\frac{dr_c}{dz}
\end{equation}\\
It is noted that effects due to dust extinction and the angular resolution of JWST's cameras are ignored for this calculation. For the field of view calculation, we use the FoV of the NIRCam instrument (9.7 arcmin$^2$).
\end{document}